\def\nnb{\\ \nonumber \\}
\def\nb{\nonumber &}
\def\nk{n_{\rm b}}

\def\rfr#1{eq. (\ref{#1})}

\def\dert#1#2{\frac{{{d}}{#1}}{{{d}}{#2}}}

\def\virg#1{``#1''}
\def\de{\right.}
\def\si{\left.}

\def\eqi{\begin{equation}}
\def\eqf{\end{equation}}
\def\eqia{\begin{eqnarray}}
\def\eqfa{\end{eqnarray}}

\def\rp#1#2{{#1\over#2}}
\def\lb#1{\label{#1}}

\def\ee{e^2}

\def\ton#1{\left(#1\right)}
\def\qua#1{\left[#1\right]}
\def\grf#1{\left\{#1\right\}}
\def\ang#1{\left\langle #1\right\rangle}

\documentclass[galaxies,article,accept,oneauthor,dvipdfm,12pt,a4paper]{mdpi} 

\lastpage{x}
\doinum{10.3390/------}
\pubvolume{xx}
\pubyear{2013}
\history{Received: xx / Accepted: xx / Published: xx}
\usepackage{amsmath,amssymb,amsthm,amscd,latexsym}
\usepackage{hyperref}
\usepackage[toc,title,titletoc]{appendix}
\usepackage{graphicx,epsfig}
\RequirePackage{color}
\allowdisplaybreaks 

\Title{Exact expressions for the pericenter precession caused by some Dark Matter distributions and constraints on them from orbital motions in the Solar System, in the double pulsar and in the Galactic center}

\Author{Lorenzo Iorio $^{1,}$*}

\address{%
$^{1}$ Italian Ministry of Education, University and Research (M.I.U.R.), Fellow of the Royal Astronomical Society (F.R.A.S.), Viale Unit\`{a} di Italia 68, 70125, Bari (BA), Italy. Tel. +39 3292399167}

\corres{lorenzo.iorio@libero.it}

\abstract{
We analytically calculate the secular precession of the pericenter of a test particle orbiting a central body surrounded by a continuous distribution of Dark Matter (DM) by using some commonly adopted \textcolor{black}{spherically symmetric} density profiles for it. We obtain exact expressions without resorting to a-priori simplifying assumptions on the orbital geometry of the test particle. Our formulas allow us to put constraints on the parameters of the DM distributions considered in several local astronomical and astrophysical scenarios, such as the Sun's planetary system, the double pulsar, and the stellar system around the supermassive black hole in Sgr A$^{\ast}$, all characterized by a wide variety of orbital configurations. As far as our Solar System is concerned, latest determinations of the supplementary perihelion precessions $\Delta\dot\varpi$ with the EPM2011 ephemerides and the common power-law DM density profile $\rho_{\rm DM}(r)=\rho_0 r^{-\gamma}\lambda^{\gamma}$ yield $5\times 10^3\ {\rm GeV\ cm^{-3}}\ (\gamma=0)\leq \rho_0\leq 8\times 10^3$ GeV cm$^{-3}\ (\gamma=4)$, corresponding to
$8.9\times 10^{-21}\ {\rm g\ cm^{-3}} \leq \rho_0\leq 1.4\times 10^{-20}$ g cm$^{-3}$,  at the Saturn's distance. From the periastron of the pulsar  PSR J0737-3039A and the same power-low DM density, one has $1.7\times 10^{16}\ {\rm GeV\ cm^{-3}}\ (\gamma=0)\leq \rho_0\leq 2\times 10^{16}\ (\gamma=4)$ GeV cm$^{-3}$, corresponding to $3.0\times 10^{-8}\ {\rm g\ cm^{-3}} \leq \rho_0\leq 3.6\times 10^{-8}$ g cm$^{-3}$. The perinigricon of the S0-2 star in Sgr A$^{\ast}$ and the power-law DM model give $1.2\times 10^{13}\ {\rm GeV\ cm^{-3}}\ (\gamma=0)\leq \rho_0\leq 1\times 10^{16}\ (\gamma=4,\ \lambda=r_{\rm min})$ GeV cm$^{-3}$, corresponding to $2.1\times 10^{-11}\ {\rm g\ cm^{-3}} \leq \rho_0\leq 1.8\times 10^{-8}$ g cm$^{-3}$.
}



\begin{document}

\section{Introduction}\lb{Introduzione}

Latest results \citep{2013arXiv1303.5076P} from the ESA's Planck mission \citep{2011A&A...536A...1P} have yielded a significative revision of the currently accepted content of non-baryonic Dark Matter (DM) of the Universe, which would now amount to about  $26.5\pm 1.1\%$  with respect to the previously accepted value of $23.2\pm 1.7\%$ from WMAP data. Indeed, the Planck-based cold DM density, normalized to the critical density, is \citep{2013arXiv1303.5076P} $\Omega_{\rm c}h^2=0.1199\pm 0.0027$, with $h=0.673\pm 0.012$, while the analysis of nine years of WMAP data yielded  \citep{2012arXiv1212.5226H} $\Omega_{\rm c}h^2=0.1138\pm 0.0045$, with $h=0.70\pm 0.022$.

Moving to  a galactic scale, the Large Area Telescope (LAT) on the Fermi Gamma Ray Space Telescope spacecraft recently discovered a gamma-ray excess at the Galactic center \cite{2012JCAP...07..054B,2012JCAP...08..007W,2012JCAP...09..032T} which may be due to DM annihilation phenomena.
The center of the Milky Way is one of the Galactic regions where a high DM density is expected. As a consequence, if DM was made of self-annihilating constituents into the Standard Model (SM) particles, signals of DM annihilations should come primarily from the Galactic center. The gamma-ray excess signature recorded by the Fermi-LAT instrument \cite{2012JCAP...07..054B,2012JCAP...08..007W,2012JCAP...09..032T} is in the form of a monochromatic gamma-ray line with an energy of about 130 GeV. Later, a second peak at 110 GeV was discovered \cite{2012JCAP...11..000T}; the same double peak spectrum was independently observed also in gamma-ray excess from nearby galaxy clusters \cite{2013ApJ...762L..22H}.
The occurrence of a double peak is a generic prediction of DM annihilation pattern in gauge theories.

There is a lingering debate concerning the amount of DM in our Solar System
\citep{2010A&A...523A..83S,2012ApJ...751...30M,2012ApJ...756...89B,POS}.
Recent, controversial estimates by Moni Bidin et al. \citep{2012ApJ...751...30M} of the DM local density in the neighbourhood of our Solar System, based on certain assumptions about the vertical velocity dispersion of old tracer stars of the thick disk at the Sun's neighborhood,  point towards a value of

\eqi\rho_{\rm DM} < 0.04\ {\rm GeV\ cm^{-3}}.\lb{monib}\eqf The bound of \rfr{monib} is smaller by about one order of magnitude than the usually accepted estimate

\eqi \rho_{\rm DM} \approx 0.3 \ {\rm GeV\ cm^{-3}}.\lb{usual} \eqf The figure in \rfr{usual}  has been considered as still valid in other recent studies \citep{2012ApJ...756...89B,POS} criticizing Moni Bidin et al. \citep{2012ApJ...751...30M}. Indeed, Bovy and Tremaine \citep{2012ApJ...756...89B}, who retained the assumptions by Moni Bidin et al. \citep{2012ApJ...751...30M} implausible, obtained

\eqi \rho_{\rm DM} = 0.3\pm 0.1 \ {\rm GeV\ cm^{-3}} \eqf by using a different hypothesis concerning the mean azimuthal velocity of the stellar tracers, while Nesti and Salucci \citep{POS} inferred

\eqi \rho_{\rm DM} = 0.43\pm{0.11}\pm{0.10} \ {\rm GeV\ cm^{-3}} \eqf
with a different method, independently of the Galactic DM density profile.
An accurate knowledge of the DM distribution in the neighbourhood of the Solar System is relevant for the attempts aimed to directly detect DM particles in laboratory-based experiments such as, e.g., CDMSI \citep{2003PhRvD..68h2002A}, CDMSII \citep{2010Sci...327.1619C}, DAMA/NaI \citep{2003NCimR..26a...1B} and its successor DAMA/LIBRA \citep{2008NIMPA.592..297B}, XENON10 \citep{2009PhRvD..80k5005A} and ZEPLIN III \citep{2005NewAR..49..277S}. Interestingly,  the silicon (Si) detectors of the CDMS II experiment has recently revealed three WIMP-candidate events at a $3\sigma$ level. \citep{2013arXiv1304.4279C}.
Moving to the Earth's neighbourhood, hints of DM might have been detected \citep{2013PhyOJ...6...40C} by the Alpha Magnetic Spectrometer (AMS) on the International Space Station, which  has recently measured an anomalous high-energy positron excess in Earth-bound cosmic rays \citep{2013PhRvL.110n1102A}.

Thus, it has become even more important to devise independent means to effectively gain information on the constitution and the distribution of such a hypothesized, elusive ingredient of the natural world.
In Section \ref{calcolo}, we will analytically calculate the secular perturbations to the  motion of a test particle orbiting a central body surrounded by a continuous DM distribution. We will adopt some \textcolor{black}{spherically symmetric} local DM density profiles \citep{2006AJ....132.2685M}
by computing the corresponding corrections $U_{\rm DM}$ to the Newtonian two-body potential through the Poisson equation.
%
%
%
%
%
%
Then, we will evaluate them onto the unperturbed Keplerian ellipse of the test particle, and we will average them  over one orbital period $P_{\rm b}$ of the test particle in order to use the standard Lagrange planetary equations \citep{befa} for the variation of the  Keplerian orbital elements. We will, thus, be able to straightforwardly infer exact expressions of the orbital rates of change due to the different DM distributions considered without resorting to any a-priori simplifying assumptions on the orbital configuration of the test particle.
Such formulas will be useful  to constrain the DM local distributions in different astrophysical and astronomical scenarios such as  planetary systems, binaries hosting compact objects, stellar systems around supermassive black holes, etc. characterized by a wide variety of orbital geometries. Only more or less approximate expressions for the DM-induced pericenter precession exist in literature.

In Section \ref{vincoli} we will apply them to the planets of the Solar System, to the double pulsar, and to the stars at the Galactic center in Sgr A$^{\ast}$.

Section \ref{ebasta} summarizes our findings.
\section{Orbital precessions for various DM density profiles}\lb{calcolo}

So far, several authors have put dynamical constraints on the DM distribution within our Solar System from orbital motions of its major bodies
\citep{1989ApJ...342..539A,1995ApJ...448..885A,1996ApJ...456..445G,2006MNRAS.371..626S,2006JCAP...05..002I,2006IJMPD..15..615K,2007IJMPD..16.1475K,2009PhLB..671..203A,2008PhRvD..77h3005F,2010JCAP...05..018I,2010IJTP...49.2506S,2012JCAP...07..047D,Pitjevi013};
for some effects of Solar System's DM on the propagation of electromagnetic waves, see \citep{2010AdSpR..45.1007A,2012JCAP...07..047D}.

In many cases, more or less approximate expressions for the anomalous perihelion precession induced by certain \textcolor{black}{spherically symmetric} DM distributions were used, in particular by considering nearly circular orbits. In this Section, we will overcome such a restriction by calculating exact expressions which can, thus,  yield more accurate constraints in view of the increasing level of accuracy in determining the orbits of some of the major bodies of the Solar System. Moreover, our results can be used also with systems characterized by highly eccentric orbits such as extrasolar planets, binaries hosting compact objects, and stellar systems orbiting supermassive black holes lurking in galactic nuclei.
\subsection{Exponential DM density profile}

By choosing an exponentially decreasing DM profile \citep{Pitjevi013}

\eqi\rho_{\rm DM}(r)=\rho_0\exp\ton{-\rp{r}{\lambda}}\lb{espon},\eqf
where $\rho_0$ is the density at a distance $r$ equal to the characteristic scale length $\lambda$, the Poisson equation yields

\eqi U_{\rm DM}(r) = -\rp{4\pi G \rho_0}{r}\lambda^3\qua{2 - \ton{2 + \rp{r}{\lambda}}\exp\ton{-\rp{r}{\lambda}} }.\lb{Upot}\eqf
The density profile of \rfr{espon} is a particular case of the Einasto profile \citep{1965TrAlm...5...87E}, often adopted to describe DM halos in galaxies \citep{2006AJ....132.2685M}.

The effect of \rfr{Upot} on the orbital motion of a test particle moving around a central body of mass $M$ surrounded by a DM continuous distribution characterized by the density profile of \rfr{espon} can be computed perturbatively by assuming \rfr{Upot} as a small correction to the  Newtonian monopole $U_{\rm N}(r) = -GM/r$.
The average of \rfr{Upot} over one orbital revolution of the test particle turns out to be

\eqi
\ang{U_{\rm DM}}\lb{Uav} = -\rp{4\pi G \rho_0\lambda^2\exp\ton{-\rp{a}{\lambda}}}{a}\qua{2 \exp\ton{\rp{a}{\lambda}}\lambda  - \ton{a + 2\lambda}I_0\ton{x} + ae I_1\ton{x} }.
\eqf
In \rfr{Uav}, $I_0\ton{x},I_1\ton{x}$ are the modified Bessel functions of the first kind, and
$x\doteq {ae}/{\lambda},$ where $a,e$ are the semimajor axis and the eccentricity of the orbit of the test particle, respectively.

The Lagrange equation for the variation of the longitude of the pericenter $\varpi$ \citep{befa} and \rfr{Uav}
yield the following secular precession

\begin{align}
\ang{\dert\varpi t} & = \rp{4\pi G \rho_0 \lambda\sqrt{1-e^2}\exp\ton{-\rp{a}{\lambda}}\qua{-I_1\ton{x} + e I_2\ton{x} } }{ae\nk}\lb{dodt},
\end{align}
where $\nk\doteq\sqrt{GM/a^3}=2\pi/P_{\rm b}$ is the Keplerian mean motion of the test particle. Although Pitjev and Pitjeva \citep{Pitjevi013} considered the density profile of \rfr{espon}, they did not explicitly calculate the resulting pericenter precession.
The precession of \rfr{dodt} is an exact result in the sense that neither a-priori simplifying assumptions on $e$ nor on $\lambda$ were assumed. Note that \rfr{dodt} is not singular in $e$ since, in the limit $e\rightarrow 0$, it reduces to

\eqi \ang{\dert\varpi t} \rightarrow -\rp{2\pi G\rho_0\exp\ton{-\rp{a}{\lambda}}}{\nk}.\eqf
In the case of an infinite length scale, i.e. for an uniform mass density, \rfr{dodt} reduces to the known result \citep{2006JCAP...05..002I,2006MNRAS.371..626S,2006IJMPD..15..615K,Pitjevi013}

\eqi \ang{\dert\varpi t} \rightarrow  -\rp{2\pi G\rho_0\sqrt{1-\ee}}{\nk}.\eqf
\subsection{Power-law DM density profile}

By assuming a power-law DM distribution  profile

\eqi\rho_{\rm DM}(r) = \rho_0\ton{\rp{r}{\lambda}}^{-\gamma},\ \gamma > 0\lb{powerlaw}\eqf
usually used for the galactic halos and in several DM-related studies \citep{2006AJ....132.2685M,2009ApJ...692.1075G}, the Poisson equation yields

\eqi U_{\rm DM}(r)=\rp{4\pi G\rho_0}{\ton{3-\gamma}\ton{2-\gamma}\lambda^{-\gamma}}r^{2-\gamma}.\lb{Upot2}\eqf

The orbital effects of \rfr{powerlaw} have been computed more or less explicitly and at various levels of approximations in \citep{2007IJMPD..16.1475K,2007PhRvD..76f2001Z,2008PhRvD..77h3005F,2010IJTP...49.2506S,2010PAN....73.1870Z,2012JCAP...07..047D}.
Actually, also the averaged pericenter precession induced by \rfr{Upot2} can be exactly computed with the Lagrange pertubative scheme \citep{befa} without resorting to any simplifying assumptions on the eccentricity $e$ of the test particle.

It turns out to be

\eqi\ang{\dert\varpi t} = \rp{\pi G\rho_0}{\nk a^{\gamma}}\sum_{j=1}^{16} p_j\ton{e;\gamma,\lambda},\lb{preces}\eqf
where the cumbersome expressions of the coefficients $p_j\ton{e;\gamma,\lambda},\ j=1,2,\ldots 16$  are explicitly displayed in Appendix \ref{appendice}.
\textcolor{black}{It is worthwhile noticing that \rfr{preces} is not any sort of somewhat truncated series expansion, and that the coefficients in Appendix \ref{appendice} are exact in the sense that no a priori simplifying assumptions were assumed in calculating them.}

By performing a series expansion of \rfr{preces} in powers of $e$ one obtains

\begin{align}
\ang{\dert\varpi t} \lb{rate2} & \approx -\rp{\pi G\rho_0}{\nk}\ton{\rp{\lambda}{a}}^{\gamma}\grf{2  +\qua{\rp{\gamma\ton{\gamma-1}-4}{4}}\ee} + \mathcal{O}\ton{e^4}.
\end{align}
The first term in \rfr{rate2} yields a perihelion shift per orbit

\eqi\Delta\varpi = \ang{\dert{\varpi}t}\left(\rp{2\pi}{\nk}\right)\eqf in agreement with that by Fr\`{e}re et al. \citep{2008PhRvD..77h3005F}.
\section{Confrontation with the observations}\lb{vincoli}

In this Section, we will use our analytical predictions of \rfr{dodt} and \rfr{preces} to infer constraints on $\rho_0$ as functions of $\lambda$ and $\gamma$ in some astronomical and astrophysical scenarios.
\subsection{Planets of the Solar System}\lb{solsys}

By following the approach adopted in several researches on DM \citep{1996ApJ...456..445G,2006MNRAS.371..626S,2006JCAP...05..002I,2006IJMPD..15..615K,2007IJMPD..16.1475K,2009PhLB..671..203A,2008PhRvD..77h3005F,2010JCAP...05..018I,2010IJTP...49.2506S,2012JCAP...07..047D,Pitjevi013}
and on other putative non-standard dynamical effects \citep{2006MNRAS.371..626S,2011MNRAS.412.2530B} in our Solar System,
we will confront our theoretically predicted pericenter precessions of \rfr{dodt} and \rfr{preces} with the latest determinations of the admissible ranges $\Delta\dot\varpi$ for any possible anomalous perihelion precessions obtained by fitting up-to-date dynamical models to a centennial record of planetary observations of several types.  The term \virg{anomalous} refers to the standard Newtonian-Einsteinian dynamics, fully modeled in the most recent planetary dynamical theories. At present, two independent teams of astronomers are engaged in producing, among other things, such corrections $\Delta\dot\varpi$ to the planetary perihelion precessions \citep{2011CeMDA.111..363F, Pitjevi013}; here we will adopt the latest estimates by Pitjev and Pitjeva  \citep{Pitjevi013}, summarized in Table \ref{tavolaf}, which are based on the most recent version of the EMP ephemerides.
\begin{table*}[ht!]
\caption{Supplementary precessions  $\Delta\dot \varpi$ of the longitudes of the perihelion for some planets of the Solar System
 estimated by Pitjev and Pitjeva \citep{Pitjevi013} with the EPM2011 ephemerides. They processed 676804 data of various kinds covering about one century (1913-2010) by estimating more than 260 parameters. While some tracking data from Cassini were used, those from Messenger, currently orbiting Mercury, were not included in the data record. Pitjev and Pitjeva \citep{Pitjevi013}  fully modeled all standard Newtonian-Einsteinian dynamics, apart from the general relativistic Lense-Thirring effect caused by the Sun's rotation. \textcolor{black}{However, its expected magnitude from general relativity is smaller than the uncertainties quoted here for the planets used in the text (Earth, Mars, Saturn), so that it does not impact our results. }. The units are  milliarcseconds per century (mas cty$^{-1}$). The errors released $\sigma_{\Delta\dot\varpi}$, in mas cty$^{-1}$ as well, generally exceed the formal, statistical ones by several times \citep{Pitjevi013}.
}\label{tavolaf}
\centering
\bigskip
\begin{tabular}{lll}
\hline\noalign{\smallskip}
Planet &   $\Delta\dot \varpi $ (mas cty$^{-1}$) & $\sigma_{\Delta\dot\varpi}$ (mas cty$^{-1}$) \\
\noalign{\smallskip}\hline\noalign{\smallskip}
Mercury & $-2.0$ & $3.0$  \\
Venus & $2.6$ & $1.6$  \\
Earth & $0.19$ & $0.19$ \\
Mars & $-0.020$ & $0.037$  \\
Jupiter & $58.7$ & $28.3$ \\
Saturn & $-0.32$ & $0.47$  \\
\noalign{\smallskip}\hline\noalign{\smallskip}
\end{tabular}
\end{table*}
DM was not explicitly modeled in the EPM routines; thus, the supplementary precessions of Table \ref{tavolaf} are well suited to put constraints on it in a phenomenological, model-independent way. Here and in the rest of the paper, \virg{model-independent} has to be intended as independent of the nature and of the physical properties of the DM particles. On the other hand, the constraints inferred in this paper depend on how DM clusters in space, i.e. on the specific density profiles adopted.  Moreover, by assuming that the whole ranges $\Delta\dot\varpi$ in Table \ref{tavolaf} are entirely due to DM, the resulting constraints will be relatively \virg{generous}, mitigating the risk of getting unrealistically tight bounds. A complementary approach which could be followed, recommended by some researchers \citep{2012CQGra..29w5027H} when non-standard effects \textcolor{black}{such as DM} are involved, consists in explicitly modeling DM and fitting such ad-hoc modified dynamical theories to the same data set in order to estimate a dedicated solved-for parameter in a least-square way. \textcolor{black}{One could try with different DM density profiles to establish the one providing the best fit to the data. Such a strategy would be particularly meaningful in the case where non-zero extra-precessions, at a statistically significant level, should be extracted from future observations.}
\subsubsection{Constraints for the exponential density profile}\lb{esziale}

Here, we deal with the exponential density profile of \rfr{espon}.

In Figure \ref{figura1} we plot the upper bounds on $\rho_0$, in GeV cm$^{-3}$, as a function of the characteristic scale length $\lambda$ over the extension of the orbits of the Earth, Mars and Saturn. We compare \rfr{dodt} to the figures in Table \ref{tavolaf}.
\begin{figure*}
\centering
\begin{tabular}{c}
\epsfig{file=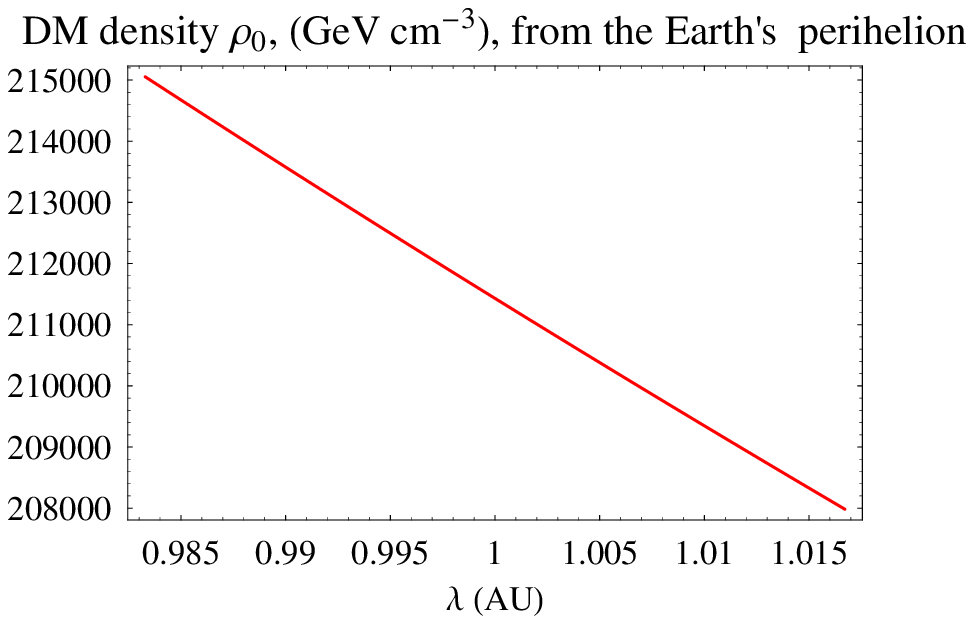,width=0.50\linewidth,clip=}\\
\epsfig{file=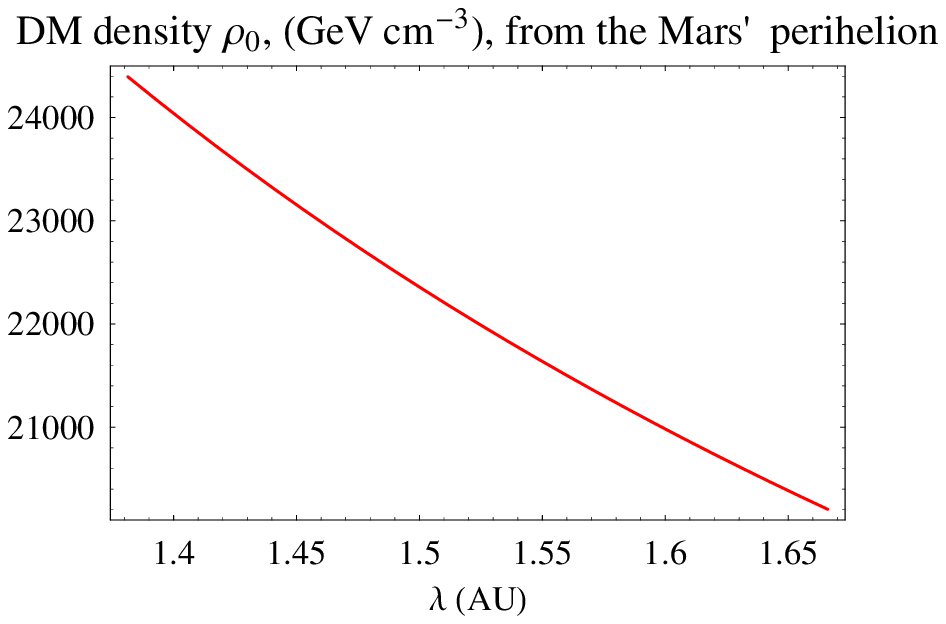,width=0.50\linewidth,clip=}\\
\epsfig{file=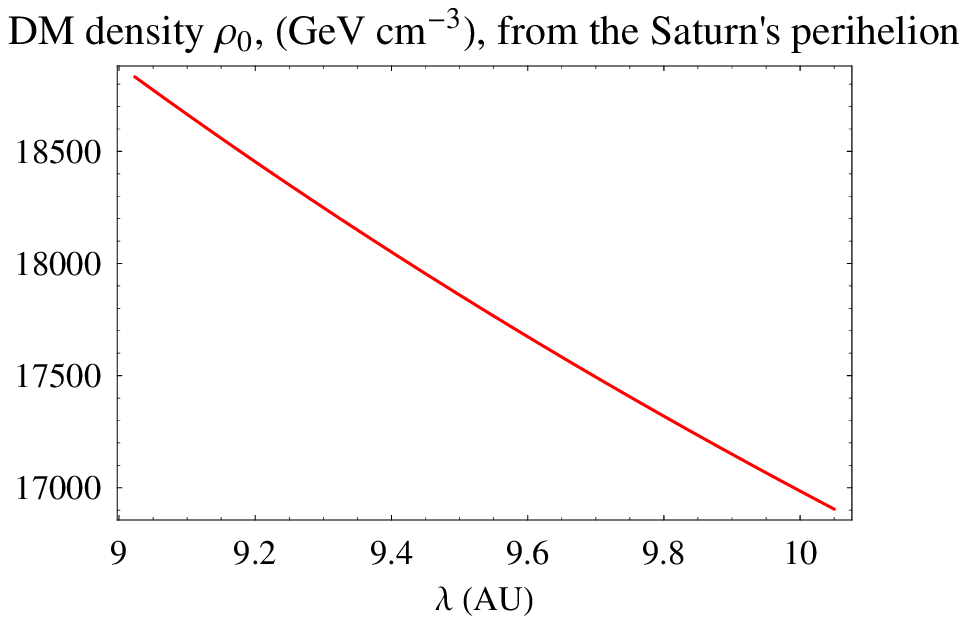,width=0.50\linewidth,clip=}\\
\end{tabular}
\caption{
Upper bounds, in GeV cm$^{-3}$ (1 Gev cm$^{-3}$ is equivalent to $1.78\times 10^{-24}$ g cm$^{-3}$), on the DM density parameter $\rho_0$ as a function of the characteristic scale length $\lambda$ of \rfr{espon}. The supplementary perihelion precessions of Table \ref{tavolaf} for Earth, Mars and Saturn were adopted along with the analytical prediction of \rfr{dodt}. The characteristic length $\lambda$, in AU, is assumed to vary from $r_{\rm min}=a(1-e)$ to $r_{\rm max}=a(1+e)$.
}\lb{figura1}

\end{figure*}
The tightest bounds come from the perihelion of Saturn; $1.7\times 10^4\ {\rm GeV\ cm^{-3}}\leq \rho_0\leq 1.9\times 10^4$ GeV cm$^{-3}$ corresponding to
$3.0\times 10^{-20}\ {\rm g\ cm^{-3}} \leq \rho_0\leq 3.4\times 10^{-20}$ g cm$^{-3}$.
\subsubsection{Constraints for the power-law density profile}

Here, the power-law distribution of \rfr{powerlaw} is considered.

Figure \ref{figura2} displays the bounds on $\rho_0$, in GeV cm$^{-3}$, as a function of the characteristic scale length $\lambda$ and of the parameter $\gamma$ for the Earth, Mars and Saturn. Also in this case, $\lambda$ covers the orbit extensions of the planets considered, while \citep{2008PhRvD..77h3005F,2012JCAP...07..047D} $0\leq \gamma\leq 4$. Also in this case, Saturn yields the tightest bounds:   $5\times 10^3\ {\rm GeV\ cm^{-3}}\ (\gamma=0)\leq \rho_0\leq 8\times 10^3$ GeV cm$^{-3}\ (\gamma=4)$ corresponding to
$8.9\times 10^{-21}\ {\rm g\ cm^{-3}} \leq \rho_0\leq 1.4\times 10^{-20}$ g cm$^{-3}$. They are about one order of magnitude better than those inferred in Section \ref{esziale}.
\begin{figure*}
\centering
\begin{tabular}{c}
\epsfig{file=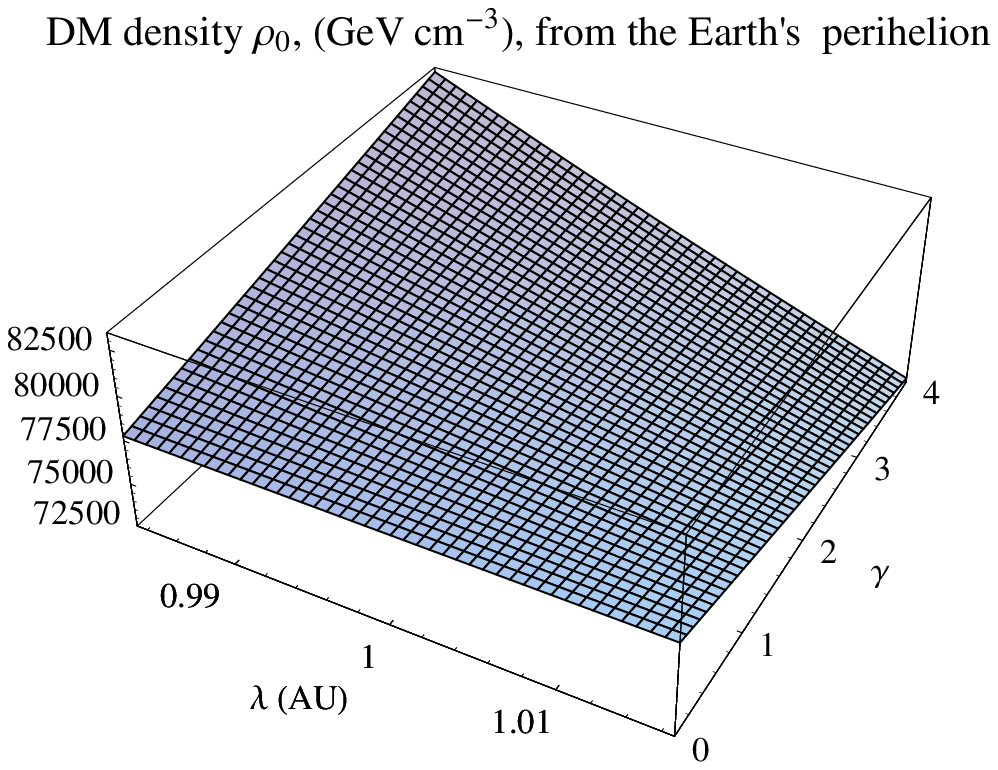,width=0.50\linewidth,clip=}\\
\epsfig{file=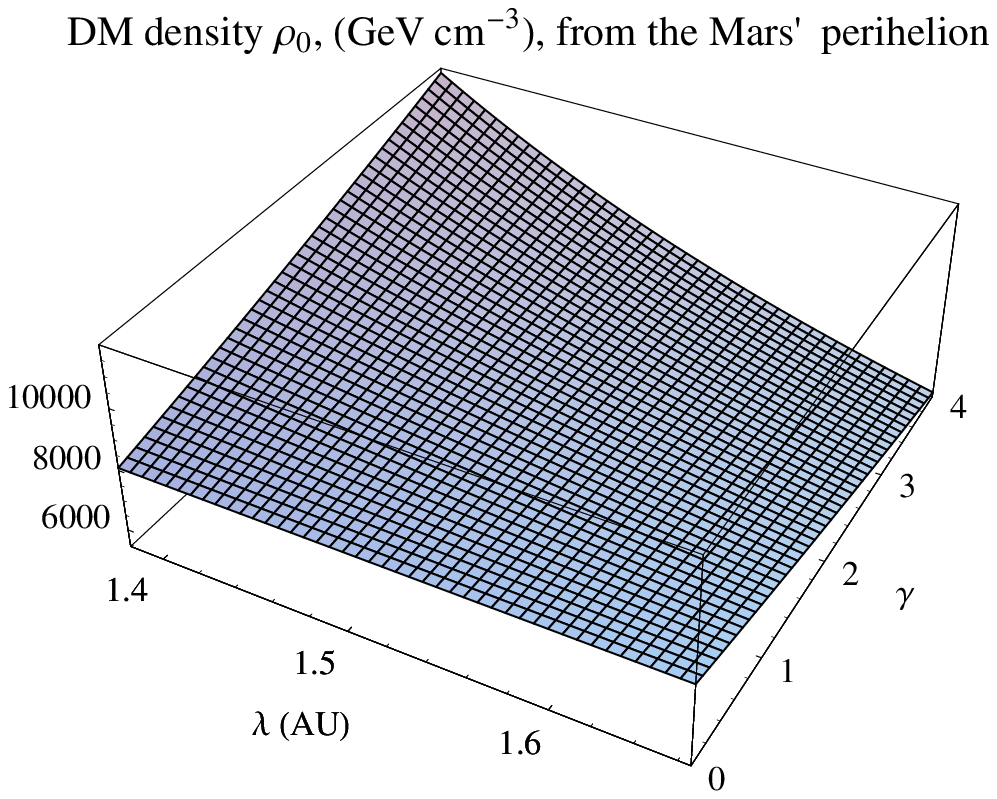,width=0.50\linewidth,clip=}\\
\epsfig{file=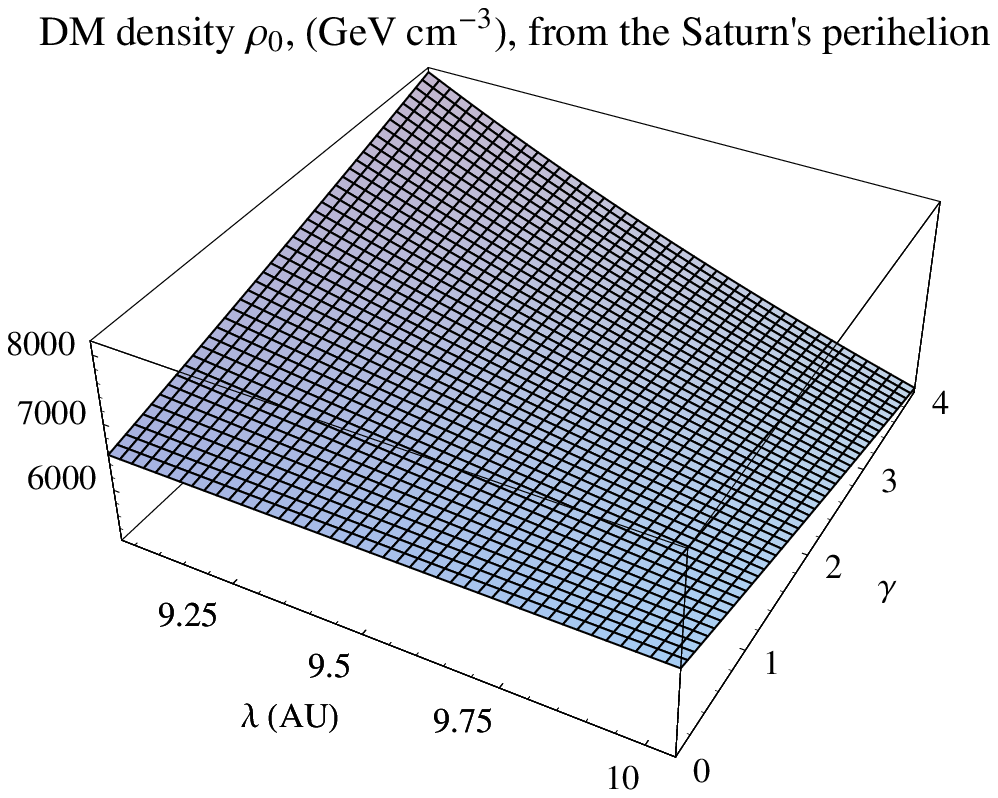,width=0.50\linewidth,clip=}\\
\end{tabular}
\caption{
Upper bounds, in GeV cm$^{-3}$ (1 Gev cm$^{-3}$ is equivalent to $1.78\times 10^{-24}$ g cm$^{-3}$), on the DM density parameter $\rho_0$ as a function of the parameters $\lambda$ and $\gamma$ of \rfr{powerlaw}. The supplementary perihelion precessions of Table \ref{tavolaf} for Earth, Mars and Saturn were adopted along with the analytical prediction of \rfr{preces}. The characteristic length $\lambda$, in AU, is assumed to vary from $r_{\rm min}=a(1-e)$ to $r_{\rm max}=a(1+e)$.
}\lb{figura2}

\end{figure*}
\subsection{The double pulsar}

 Thanks to their high matter density, astrophysical compact objects such as neutron stars should be able to efficiently capture significative amounts of DM \citep{1989PhRvD..40.3221G,1990PhLB..238..337G,1999ApJ...512..282L,2008PhRvD..77d3515B,2008PhRvD..77b3006K, 2010PhRvD..82f3531K,2010PhRvD..81l3521D,2010A&A...522A..16G,2011PhLB..695...19C,2011ApJS..197...37B,2012JCAP...01..023Y} depending on the nature of its particles, on how it clusters in space and on the history of the stars themselves.  Concentrations of DM around neutron stars may give rise to phenomena such as gamma-ray emission \cite{2011ApJS..197...37B,2012JCAP...01..023Y} which, on the one hand, depend on the physical properties of DM particles themselves \cite{1985PhR...117...75H} and, on the other hand, can be generated also by other physical mechanisms \cite{2013A&A...551A..17Z,2013arXiv1303.7352G}. Thus, it is valuable to constrain the DM density around binaries hosting neutron stars in a phenomenological and model-independent way.

To this aim, we look at the double pulsar PSR J0737-3039A/B \citep{2003Natur.426..531B,2004Sci...303.1153L}.
As in Section \ref{solsys}, we consider both \rfr{espon} and \rfr{powerlaw}. In this case, we use the periastron precession of PSR J0737-3039A, which is nowadays determined with an accuracy of $6.8\times 10^{-4}$ deg yr$^{-1}$ from timing measurements \citep{2006Sci...314...97K}. Nonetheless, care is required in straightforwardly using such a figure in comparisons with theoretical predictions of non-standard effects to infer constraints on them. Indeed, a larger uncertainty should, actually, be considered on the periastron rate of PSR J0737-3039A because of the mismodeling in its
1PN periastron precession \citep{2009NewA...14..196I}; it may be as large as $0.03$ deg yr$^{-1}$ \citep{2009NewA...14..196I}.
\begin{figure*}
\centering
\begin{tabular}{c}
\epsfig{file=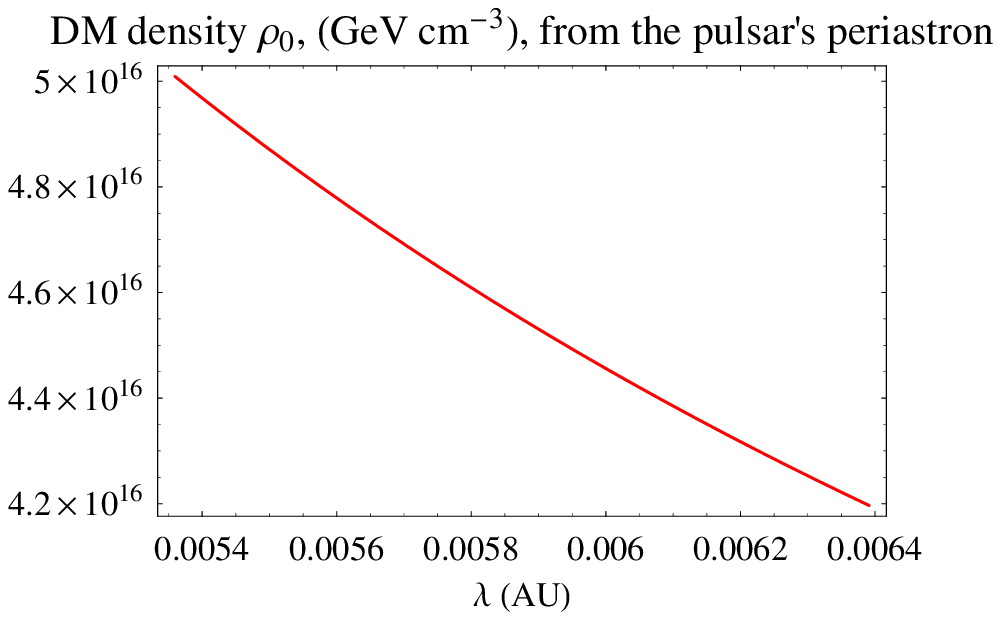,width=0.50\linewidth,clip=}\\
\epsfig{file=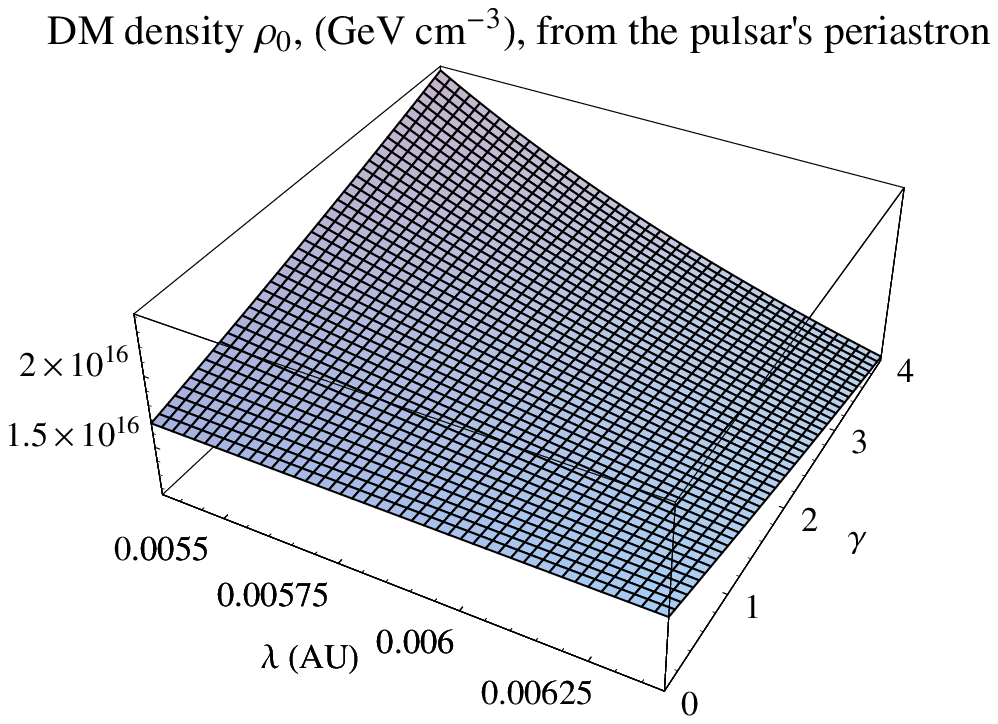,width=0.50\linewidth,clip=}\\
\end{tabular}
\caption{
Upper bounds, in GeV cm$^{-3}$ (1 Gev cm$^{-3}$ is equivalent to $1.78\times 10^{-24}$ g cm$^{-3}$), on the DM density parameter $\rho_0$ in the PSR J0737-3039A/B system as a function of $\lambda$ (top panel) of \rfr{espon}, and of  $\lambda$ and $\gamma$ of \rfr{powerlaw} (bottom panel).  The characteristic length $\lambda$, in AU, is assumed to vary from $r_{\rm min}=a(1-e)$ to $r_{\rm max}=a(1+e)$. A conservative periastron rate uncertainty of $0.03$ deg yr$^{-1}$ \citep{2009NewA...14..196I} was assumed for PSR J0737-3039A. It was compared to the theoretical predictions of \rfr{dodt} (top panel) and of \rfr{preces} (bottom panel).
}\lb{figurapsr}

\end{figure*}
In Figure \ref{figurapsr} the bounds for both \rfr{espon} and \rfr{powerlaw} are shown.  The power-law profile provides the tightest constraints. Indeed, over the extension of the double pulsar system, it yields  $1.7\times 10^{16}\ {\rm GeV\ cm^{-3}}\ (\gamma=0)\leq \rho_0\leq 2\times 10^{16}\ (\gamma=4)$ GeV cm$^{-3}$ corresponding to
$3.0\times 10^{-8}\ {\rm g\ cm^{-3}} \leq \rho_0\leq 3.6\times 10^{-8}$ g cm$^{-3}$, while the bounds from the exponential density profile are of the order of $5\times 10^{16}$ GeV cm$^{-3}$ corresponding to $9\times 10^{-8}$ g cm$^{-3}$.
\subsection{The stellar system orbiting the Galactic  black hole in Sgr A$^\ast$}

The issue of the DM distribution at the center of galaxies in presence of supermassive black holes has been treated in, e.g., \citep{1999PhRvL..83.1719G,2006PhRvD..74f3511H,2007PhRvD..76f2001Z,2010arXiv1001.3706M,2010PAN....73.1870Z,2012APS..APR.X8003S}. It depends on several factors such as \citep{1999PhRvL..83.1719G} the galactic halo density profile itself \citep{2013ApJ...765...10K} and the nature of the DM particles themselves. Moreover, as pointed out in   \citep{2012APS..APR.X8003S}, significant differences in the final DM distribution close to the black hole are found depending on the theoretical scheme adopted for the calculation.

So far, the presence of DM at the center of the Milky Way has been indirectly guessed from certain phenomena such as gamma-ray emissions interpreted as DM annihilation \citep{2012PhRvD..86h3516B,2013PhRvD..87d3516L}. However, such interpretations are not free from more or less tested assumptions about several other concurring physical phenomena \citep{2012CRPhy..13..740L,2012PDU.....1..194B}. Recently, the Fermi satellite detected a gamma-ray excess at the Galactic center which may be due to DM annihilation phenomena \cite{2012JCAP...07..054B,2012JCAP...08..007W,2012JCAP...09..032T}. Thus, it is important to constrain DM distributions at the Galactic center in a dynamical, model-independent way.	

To this aim, we will look at the orbital motions of the stellar system \citep{2009ApJ...692.1075G,2009ApJ...707L.114G,2012Sci...338...84M} revolving about the supermassive black hole located at the Galactic center in Sgr A$^{\ast}$ \citep{2007gsbh.book.....M}. So far, available data cover one full orbital revolution for two stars: S$0$-$2$, having an orbital period $P_{\rm b} = 16.17\pm 0.15$ yr and eccentricity $e = 0.898\pm 0.005$ \citep{2012Sci...338...84M}, and S$0$-$102$, characterized by $P_{\rm b} = 11.5\pm 0.3$ yr and eccentricity $e = 0.68\pm 0.02$ \citep{2012Sci...338...84M}. The mass of the black hole is $M_\bullet=(4.1\pm 0.4)\times 10^6 M_{\odot}$ \citep{2012Sci...338...84M}.
From such figures, a naive, order-of-magnitude evaluation on the accuracy which could be reached in determining the stellar perinigricon precessions can be made; for S0-2 we have $\sigma_{\dot\omega}\approx 0.6$ deg yr$^{-1}$, while for S0-102 it can be inferred $\sigma_{\dot\omega}\approx 4.8$ deg yr$^{-1}$. Then, we will adopt S0-2 as probe to constrain DM density as done in the previous Sections. In view of its high eccentricity, the exact formulas of \rfr{dodt} and \rfr{preces} are well suited for such a task.
\begin{figure*}
\centering
\begin{tabular}{c}
\epsfig{file=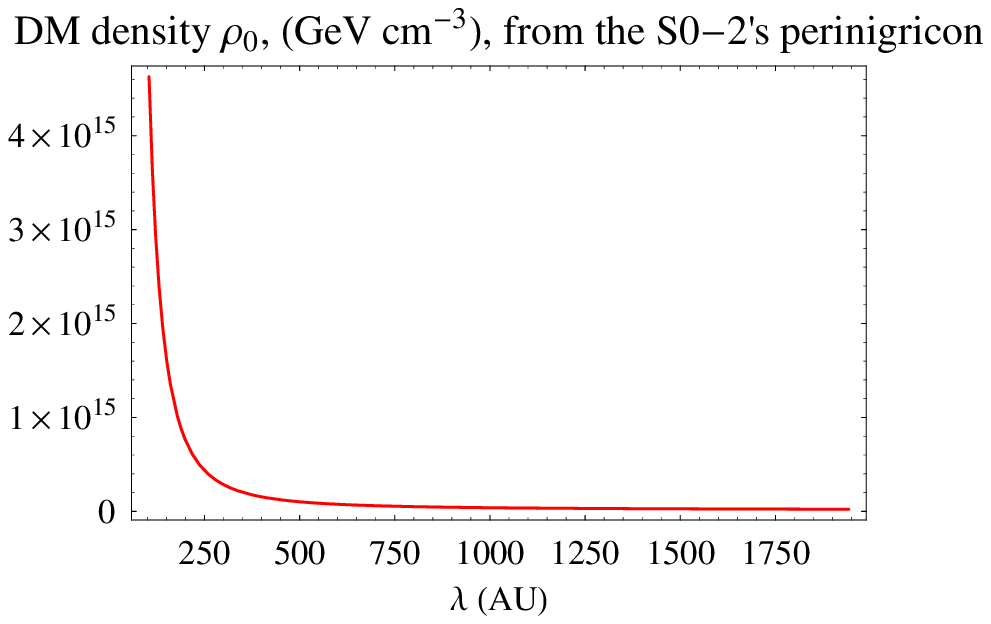,width=0.50\linewidth,clip=}\\
\epsfig{file=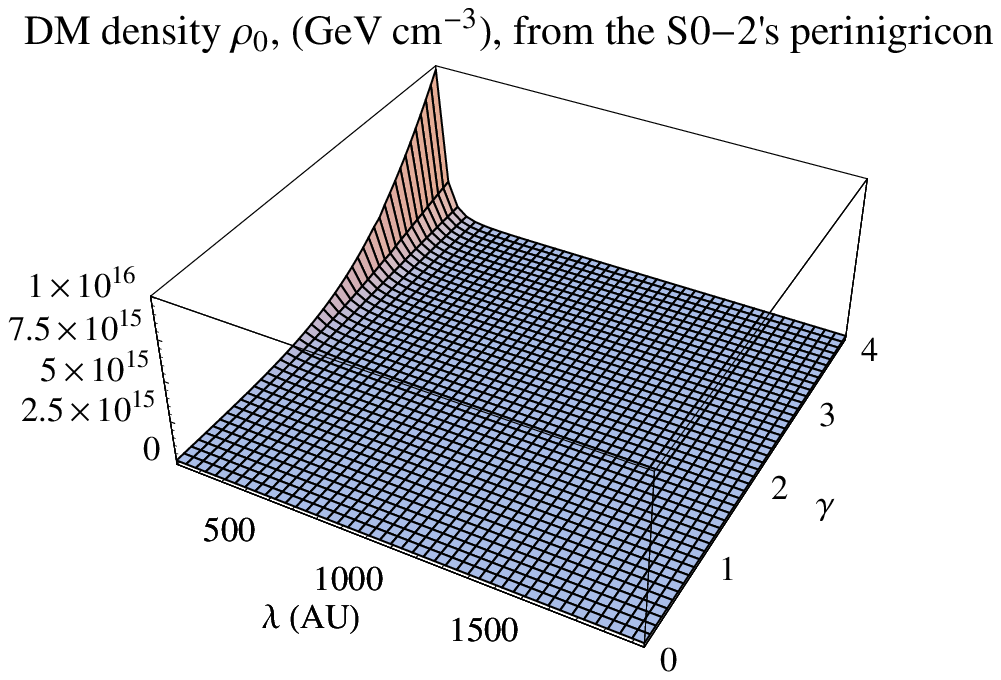,width=0.50\linewidth,clip=}\\
\end{tabular}
\caption{
Upper bounds, in GeV cm$^{-3}$ (1 Gev cm$^{-3}$ is equivalent to $1.78\times 10^{-24}$ g cm$^{-3}$), on the DM density parameter $\rho_0$ in the Sgr A$^{\ast}$$-$S0-2 system as a function of $\lambda$ (top panel) of \rfr{espon}, and of  $\lambda$ and $\gamma$ of \rfr{powerlaw} (bottom panel).  The characteristic length $\lambda$, in AU, is assumed to vary from $r_{\rm min}=a(1-e)$ to $r_{\rm max}=a(1+e)$. A conservative perinigricon rate uncertainty of $0.6$ deg yr$^{-1}$  was assumed for S0-2. It was compared to the theoretical predictions of \rfr{dodt} (top panel) and of \rfr{preces} (bottom panel).
}\lb{figuraS2}

\end{figure*}
From Figure \ref{figuraS2} it turns out that the bounds for the exponential density vary sensibly over the S0-2 orbit extension in view of its high eccentricity; indeed, it is $\rho_0\approx 4\times 10^{15}$ GeV cm$^{-3}$, corresponding to $7\times 10^{-9}$ g cm$^{-3}$, for $\lambda=r_{\rm min}$, while it reduces to $\rho_0\approx 2\times 10^{13}$ GeV cm$^{-3}$, corresponding to $3\times 10^{-11}$ g cm$^{-3}$, for $\lambda = r_{\rm max}$.
As pointed out in \citep{2009ApJ...692.1075G}, as far as supposedly baryonic DM is concerned, both theoretical \citep{1977ApJ...216..883B,1980ApJ...242.1232Y} and observational \citep{2003ApJ...594..812G} motivations for considering a power-law profile such as \rfr{powerlaw} at the galactic centers exist in literature. The same kind of potential was considered for non-baryonic DM as well \cite{2006PhRvD..74f3511H,2010PAN....73.1870Z}. According to Figure \ref{figuraS2}, it yields  $1.2\times 10^{13}\ {\rm GeV\ cm^{-3}}\ (\gamma=0)\leq \rho_0\leq 1\times 10^{16}\ (\gamma=4,\ \lambda=r_{\rm min})$ GeV cm$^{-3}$ corresponding to
$2.1\times 10^{-11}\ {\rm g\ cm^{-3}} \leq \rho_0\leq 1.8\times 10^{-8}$ g cm$^{-3}$.

For an earlier idea of constraining the  DM concentration near the Galactic center from the perinigricon of S0-2, see \citep{2007PhRvD..76f2001Z,2010PAN....73.1870Z}.
\section{Summary and conclusions}\lb{ebasta}

Recent results concerning the amount of non-baryonic DM, both at global and local scales,
stress the need of further deepening the research of accurate and independent strategies to gain information about the distribution of such a hypothetical key ingredient of the natural world.

To this aim, we looked at the effects induced by some spherically symmetric DM density profiles on the  motion of a test particle orbiting a localized body surrounded by an extended DM distribution. In view of the increasing accuracy in the determination of the orbits of the planets of our Solar System and of the possibility of looking also at different astronomical and astrophysical laboratories characterized by a wide variety of orbital configurations, we analytically calculated the DM-induced pericenter precession without resorting to any a-priori simplifying assumptions concerning the orbital geometry of the test particle.
In this respect, our results, obtained perturbatively by means of the Lagrange planetary equations, are exact, being valid for any value of the eccentricity of the orbit of the test particle.

We considered an exponentially decreasing profile $\rho_{\rm DM}(r)=\rho_0\exp\ton{-r/\lambda}$ and a standard power-law model $\rho_{\rm DM}(r)=\rho_0 r^{-\gamma}\lambda^{\gamma}$. We compared our analytical predictions with the latest observational determinations for some planets of our Solar System obtained with the EPM2011 ephemerides, the double pulsar, and the S0-2 star orbiting the supermassive black hole in Sgr A$^{\ast}$. The tightest constraints, obtained for the power-law model, came from the supplementary precessions of the planetary perihelia. We obtained  $5\times 10^3\ {\rm GeV\ cm^{-3}}\ (\gamma=0)\leq \rho_0\leq 8\times 10^3$ GeV cm$^{-3}\ (\gamma=4)$, corresponding to
$8.9\times 10^{-21}\ {\rm g\ cm^{-3}} \leq \rho_0\leq 1.4\times 10^{-20}$ g cm$^{-3}$,  at the Saturn's distance. From the periastron of PSR J0737-3039A  we inferred  $1.7\times 10^{16}\ {\rm GeV\ cm^{-3}}\ (\gamma=0)\leq \rho_0\leq 2\times 10^{16}\ (\gamma=4)$ GeV cm$^{-3}$, corresponding to $3.0\times 10^{-8}\ {\rm g\ cm^{-3}} \leq \rho_0\leq 3.6\times 10^{-8}$ g cm$^{-3}$. The perinigricon of the S0-2 star in Sgr A$^{\ast}$  gave $1.2\times 10^{13}\ {\rm GeV\ cm^{-3}}\ (\gamma=0)\leq \rho_0\leq 1\times 10^{16}\ (\gamma=4,\ \lambda=r_{\rm min})$ GeV cm$^{-3}$, corresponding to $2.1\times 10^{-11}\ {\rm g\ cm^{-3}} \leq \rho_0\leq 1.8\times 10^{-8}$ g cm$^{-3}$.

Our results can be used in future when new, more accurate data will be collected and processed. As a complementary approach which could be followed, DM dynamical effects should be explicitly modeled in the softwares used to reduce the planetary observations, and dedicated solved-for parameters should be estimated in fitting the newly constructed models of the forces acting on the planets to given data records.
\renewcommand\appendix{\par
\setcounter{section}{0}%
\setcounter{subsection}{0}%
\setcounter{table}{0}
\setcounter{figure}{0}
\setcounter{equation}{0}
\gdef\thetable{\Alph{table}}
\gdef\thefigure{\Alph{figure}}
\gdef\theequation{\Alph{section}.\arabic{equation}}
\section*{Appendix}
\gdef\thesection{\Alph{section}}
\setcounter{section}{0}}

\appendix
\section{The coefficients of the pericenter precession due to the power-law density}\lb{appendice}
Below, the exact expressions of the coefficients $p_j\ton{e,;\gamma,\lambda},\ j=1,2,\ldots,16$ of the pericenter precession of \rfr{preces}, induced by \rfr{powerlaw}-\rfr{Upot2}, are listed. No simplifying assumptions on the eccentricity $e$ were assumed in computing them. They contain the hypergeometric function $_2 F_1\ton{q,w;b;d}$ \citep{1972hmfw.book.....A}.

\begin{align}
p_1 \lb{p1}\nb = -\rp{2  \ton{1+e}^{-\gamma}\lambda^{\gamma}}{15\ton{-1+e}^2 e\sqrt{1-\ee}\ton{6-5\gamma+\gamma^2}^2 }\grf{\de\nnb
\nb \si (\gamma -3) (\gamma  (\gamma  (16 \gamma  (8 \gamma -61)+2687)-3204)+1405) e^6
-\de\nnb
\nb -\si (\gamma -2) (\gamma  (16 \gamma  (\gamma  (8 \gamma -57)+140)-2265)+815) e^5
+\de\nnb
\nb +\si (\gamma  (\gamma(\gamma  (464 \gamma -3917)+11838)-15306)+7155) e^4
+\de\nnb
\nb +\si 2 (\gamma  (\gamma  (24 \gamma  (5 \gamma -38)+2813)-4095)+2230) e^3
+\de\nnb
\nb +\si (\gamma  ((5099-915 \gamma ) \gamma -8991)+5095)e^2
+\de\nnb
\nb + \si 15 (\gamma  (35 \gamma -151)+154) e
-\de\nnb
\nb -\si 5 (3 \gamma  ((\gamma -8) \gamma +24)-73)\de\nnb
& \si}\ _2 F_1\ton{\rp{1}{2},\gamma;1;\rp{2e}{1+e}},\nnb
p_2 \lb{p2} \nb = -\rp{2  \ton{1+e}^{-\gamma}\lambda^{\gamma}}{15\ton{-1 + e}^3 e\sqrt{1-\ee}\ton{6-5\gamma+\gamma^2}^2 }\grf{\de\nnb
\nb \si 2 (\gamma -3) (\gamma -2) (\gamma  (2 \gamma  (8 \gamma  (8 \gamma -55)+1093)-2421)+985) e^7- \de\nnb
\nb -\si (\gamma -2) (\gamma  (4 \gamma  (\gamma  (8 \gamma  (8 \gamma -73)+2065)-3593)+12571)-4325) e^6+ \de\nnb
\nb + \si (\gamma  (\gamma  (2 \gamma  (12 \gamma  (52 \gamma -485)+21371)-77699)+69856)-24575) e^5+ \de\nnb
\nb + \si (\gamma  (8 \gamma  (5 \gamma  (3 \gamma  (4\gamma -47)+616)-6337)+49979)-19385) e^4 - \de\nnb
\nb -\si 2 (\gamma  (\gamma  (\gamma  (990 \gamma -7771)+21801)-26063)+11335) e^3+ \de\nnb
\nb +\si (\gamma  (\gamma  (60 (35-3 \gamma ) \gamma -8891)+15549)-9320) e^2+ \de\nnb
\nb +\si 5 (3 \gamma -5) (\gamma  (30 \gamma -119)+95) e- \de\nnb
\nb -\si 5 (9 \gamma  (2 \gamma -9)+89) \de\nnb
& \si}\ _2 F_1\ton{\rp{1}{2},\gamma;2;\rp{2e}{1+e}},\nnb
p_3 \lb{p3} \nb = -\rp{2   \gamma \ton{1+e}^{-\gamma}\lambda^{\gamma} }{ 5\ton{-1 + e} \sqrt{1-\ee}\ton{6-5\gamma+\gamma^2}^2 }\grf{\de\nnb
\nb \si (4 \gamma  (\gamma  (8 \gamma -53)+104)-225) e^3+\de\nnb
\nb +\si  2 (4 \gamma -11) (4 \gamma -5) e^2-\de\nnb
\nb -\si 5 (4 \gamma  (3 \gamma -14)+63) e+\de\nnb
\nb +\si 10 \de\nnb
& \si}\ _2 F_1\ton{\rp{1}{2},1+\gamma;2;\rp{2e}{1+e}},\nnb
p_4 \lb{p4} \nb = -\rp{   \gamma \ton{1+e}^{-\gamma}\lambda^{\gamma} }{ 5\ton{-1 + e}^2 \sqrt{1-\ee}\ton{6-5\gamma+\gamma^2}^2 }\grf{\de\nnb
\nb \si (\gamma  (\gamma  (8 \gamma  (8 \gamma -67)+1669)-2256)+1070) e^4+\de\nnb
\nb +\si(\gamma  (4 \gamma  (4 \gamma -45)+489)-335) e^3-\de\nnb
\nb -\si (2 \gamma -3) (3 \gamma -10) (20 \gamma -47) e^2-\de\nnb
\nb -\si 5 (\gamma(12 \gamma -55)+65) e+\de\nnb
\nb +\si 5 (3 \gamma -8) (3 \gamma -7)\de\nnb
& \si}\ _2 F_1\ton{\rp{1}{2},1+\gamma;3;\rp{2e}{1+e}},\nnb
p_5 \lb{p5} \nb = -\rp{2   \ton{1+e}^{1-\gamma}\lambda^{\gamma} }{ 5\ton{-1 + e}^2 e\sqrt{1-\ee}\ton{6-5\gamma+\gamma^2}^2 }\grf{\de\nnb
\nb \si (\gamma -3) (4 \gamma  (\gamma  (8 \gamma -53)+104)-225) e^5+\de\nnb
\nb +\si (\gamma  (1839-4 \gamma  (\gamma  (8 \gamma -85)+311))-895) e^4+\de\nnb
\nb +\si (\gamma  (4 \gamma  (9 \gamma -71)+575)-255) e^3+\de\nnb
\nb +\si (2\gamma -5) (\gamma  (30 \gamma -107)+123) e^2-\de\nnb
\nb -\si 10 (\gamma  (12 \gamma -55)+60) e+\de\nnb
\nb +\si 10\de\nnb
& \si}\ _2 F_1\ton{-\rp{1}{2},\gamma;1;\rp{2e}{1+e}},\nnb
p_6 \lb{p6} \nb = -\rp{2   \ton{1+e}^{1-\gamma}\lambda^{\gamma} }{ 5\ton{-1 + e}^3 e\sqrt{1-\ee}\ton{6-5\gamma+\gamma^2}^2 }\grf{\de\nnb
\nb\si (\gamma -3) (\gamma  (\gamma  (8 \gamma  (8 \gamma -67)+1669)-2256)+1070) e^6+\de\nnb
\nb +\si (\gamma  (\gamma  (\gamma  (8 (101-8 \gamma ) \gamma -4025)+9781)-11407)+4950) e^5+\de\nnb
\nb +\si (\gamma(\gamma  (2 \gamma  (92 \gamma -727)+4311)-5616)+2600) e^4+\de\nnb
\nb +\si \left(\gamma  \left(2 \gamma  \left(60 \gamma ^2-669 \gamma +2483\right)-7631\right)+4300\right) e^3+\de\nnb
\nb +\si (\gamma((1891-255 \gamma ) \gamma -4409)+3210) e^2-\de\nnb
\nb -\si 5 (3 \gamma  (\gamma  (3 \gamma -19)+42)-94) e+\de\nnb
\nb +\si 5 (3 \gamma -8) (3 \gamma -7)\de\nnb
& \si}\ _2 F_1\ton{-\rp{1}{2},\gamma;2;\rp{2e}{1+e}},\nnb
p_7 \lb{p7} \nb = -\rp{2   \gamma \ton{1+e}^{-\gamma}\lambda^{\gamma} }{ 15\ton{1 - e^2}^{3/2} e\ton{6-5\gamma+\gamma^2}^2 }\grf{\de\nnb
\nb\si(\gamma  (\gamma  (16 \gamma  (8 \gamma -61)+2687)-3204)+1405) e^5+\de\nnb
\nb +\si (\gamma  ((447-64 \gamma ) \gamma -939)+590) e^4-\de\nnb
\nb -\si 2 (\gamma  (\gamma  (120 \gamma -649)+1066)-525) e^3+\de\nnb
\nb +\si 10 (3\gamma -8) (7 \gamma -11) e^2+\de\nnb
\nb +\si 5 (3 (\gamma -8) \gamma +37) e+\de\nnb
\nb +\si 15 (\gamma -3) (\gamma -2)\de\nnb
& \si}\ _2 F_1\ton{\rp{3}{2},1+\gamma;2;\rp{2e}{1+e}},\nnb
p_8 \lb{p8} \nb = \rp{  \gamma  \ton{1+e}^{-{3/2}-\gamma}\lambda^{\gamma} }{ 15\ton{1 - e}^{5/2} \ton{6-5\gamma+\gamma^2}^2 }\grf{\de\nnb
\nb\si 256 e^5 \gamma ^5-\de\nnb
\nb -\si 32 e^3 (e (71 e+6)+15) \gamma ^4+\de\nnb
\nb +\si 4 e (e (e (e (1973 e+348)+870)+60)+45) \gamma ^3-\de\nnb
\nb -\si 2 (e (e (e (e (6793 e+1793)+4446)+826)+465)+45) \gamma ^2+\de\nnb
\nb +\si (e (e (e (e (11654 e+3925)+9608)+3518)+1370)+405) \gamma -\de\nnb
\nb -\si 5 (e (e (e (e (788 e+317)+764)+458)+104)+89)\de\nnb
& \si}\ _2 F_1\ton{\rp{3}{2},1+\gamma;3;\rp{2e}{1+e}},\nnb
p_9 \lb{p9} \nb = \rp{2   \ton{1-e}^{-\gamma}\lambda^{\gamma} }{ 15\ton{1 + e}^{5/2}e\sqrt{1-e} \ton{6-5\gamma+\gamma^2}^2 }\grf{\de\nnb
\nb\si(\gamma -3) (\gamma  (\gamma  (16 \gamma  (8 \gamma -61)+2687)-3204)+1405) e^6+\de\nnb
\nb +\si(\gamma -2) (\gamma  (16 \gamma  (\gamma  (8 \gamma -57)+140)-2265)+815) e^5+\de\nnb
\nb +\si(\gamma  (\gamma(\gamma  (464 \gamma -3917)+11838)-15306)+7155) e^4-\de\nnb
\nb -\si 2 (\gamma  (\gamma  (24 \gamma  (5 \gamma -38)+2813)-4095)+2230) e^3+\de\nnb
\nb +\si (\gamma  ((5099-915 \gamma ) \gamma -8991)+5095)e^2-\de\nnb
\nb -\si 15 (\gamma  (35 \gamma -151)+154) e-\de\nnb
\nb -\si 5 (3 \gamma  ((\gamma -8) \gamma +24)-73)\de\nnb
& \si}\ _2 F_1\ton{\rp{1}{2},\gamma;1;\rp{2e}{-1+e}},\nnb
p_{10} \lb{p10} \nb = \rp{2   \ton{1-e}^{-\gamma}\lambda^{\gamma} }{ 15\ton{1 + e}^{7/2}e\sqrt{1-e} \ton{6-5\gamma+\gamma^2}^2 }\grf{\de\nnb
\nb \si 2 (\gamma -3) (\gamma -2) (\gamma  (2 \gamma  (8 \gamma  (8 \gamma -55)+1093)-2421)+985) e^7+\de\nnb
\nb +\si(\gamma -2) (\gamma  (4 \gamma  (\gamma  (8 \gamma  (8 \gamma -73)+2065)-3593)+12571)-4325) e^6+\de\nnb
\nb +\si (\gamma  (\gamma  (2 \gamma  (12 \gamma  (52 \gamma -485)+21371)-77699)+69856)-24575) e^5+\de\nnb
\nb +\si (\gamma  (-8 \gamma  (5 \gamma  (3 \gamma  (4\gamma -47)+616)-6337)-49979)+19385) e^4-\de\nnb
\nb -\si 2 (\gamma  (\gamma  (\gamma  (990 \gamma -7771)+21801)-26063)+11335) e^3+\de\nnb
\nb +\si (\gamma  (\gamma  (60 \gamma  (3 \gamma -35)+8891)-15549)+9320) e^2+\de\nnb
\nb +\si 5 (3 \gamma -5) (\gamma  (30 \gamma -119)+95) e+\de\nnb
\nb\si + 5 (9 \gamma  (2 \gamma -9)+89)\de\nnb
& \si}\ _2 F_1\ton{\rp{1}{2},\gamma;2;\rp{2e}{-1+e}},\nnb
p_{11} \lb{p11} \nb = -\rp{2  \gamma \ton{1-e}^{-{1/2}-\gamma}\lambda^{\gamma} }{ 5\ton{1 + e}^{3/2} \ton{6-5\gamma+\gamma^2}^2 }\grf{\de\nnb
\nb\si 32 e^3 \gamma ^3-\de\nnb
\nb -\si 4 e (e (53 e+8)+15) \gamma ^2+\de\nnb
\nb +\si 8 e (4 e (13 e+4)+35) \gamma -\de\nnb
\nb -\si 5 (e (e (45 e+22)+63)+2)\de\nnb
& \si}\ _2 F_1\ton{\rp{1}{2},1+\gamma;2;\rp{2e}{-1+e}},\nnb
p_{12} \lb{p12} \nb = -\rp{  \gamma \ton{1-e}^{-\gamma}\lambda^{\gamma} }{ 5\ton{1 + e}^{5/2}\sqrt{1-e} \ton{6-5\gamma+\gamma^2}^2 }\grf{\de\nnb
\nb\si(\gamma  (\gamma  (8 \gamma  (8 \gamma -67)+1669)-2256)+1070) e^4+\de\nnb
\nb + \si(\gamma  (4 (45-4 \gamma ) \gamma -489)+335) e^3-\de\nnb
\nb -\si (2 \gamma -3) (3 \gamma -10) (20 \gamma -47) e^2+\de\nnb
\nb +\si 5 (\gamma(12 \gamma -55)+65) e+\de\nnb
\nb +\si 5 (3 \gamma -8) (3 \gamma -7)\de\nnb
& \si}\ _2 F_1\ton{\rp{1}{2},1+\gamma;3;\rp{2e}{-1+e}},\nnb
p_{13} \lb{p13} \nb = -\rp{2   \ton{1-e}^{1/2-\gamma}\lambda^{\gamma} }{ 5 e\ton{1 + e}^{5/2}\ton{6-5\gamma+\gamma^2}^2 }\grf{\de\nnb
\nb\si (\gamma -3) (4 \gamma  (\gamma  (8 \gamma -53)+104)-225) e^5+\de\nnb
\nb +\si (\gamma  (4 \gamma  (\gamma  (8 \gamma -85)+311)-1839)+895) e^4+\de\nnb
\nb +\si(\gamma  (4 \gamma  (9 \gamma -71)+575)-255)e^3+\de\nnb
\nb +\si (\gamma  (4 (91-15 \gamma ) \gamma -781)+615) e^2-\de\nnb
\nb -\si 10 (\gamma  (12 \gamma -55)+60) e\de\nnb
\nb -\si 10\de\nnb
& \si}\ _2 F_1\ton{-\rp{1}{2},\gamma;1;\rp{2e}{-1+e}},\nnb
p_{14} \lb{p14} \nb = -\rp{2  \ton{-1+e}^4 \ton{1-e}^{-\gamma}\lambda^{\gamma} }{ 5 e\ton{1 - e^2}^{7/2}\ton{6-5\gamma+\gamma^2}^2 }\grf{\de\nnb
\nb\si (\gamma -3) (\gamma  (\gamma  (8 \gamma  (8 \gamma -67)+1669)-2256)+1070) e^6+\de\nnb
\nb +\si(\gamma  (\gamma  (\gamma  (8 \gamma  (8 \gamma -101)+4025)-9781)+11407)-4950) e^5+\de\nnb
\nb +\si (\gamma  (\gamma(2 \gamma  (92 \gamma -727)+4311)-5616)+2600) e^4+\de\nnb
\nb +\si \left(\gamma  \left(7631-2 \gamma  \left(60 \gamma ^2-669 \gamma +2483\right)\right)-4300\right) e^3+\de\nnb
\nb +\si (\gamma  ((1891-255\gamma ) \gamma -4409)+3210) e^2+\de\nnb
\nb +\si 5 (3 \gamma  (\gamma  (3 \gamma -19)+42)-94) e+\de\nnb
\nb +\si 5 (3 \gamma -8) (3 \gamma -7)\de\nnb
& \si}\ _2 F_1\ton{-\rp{1}{2},\gamma;2;\rp{2e}{-1+e}},\nnb
p_{15} \lb{p15} \nb = -\rp{2  \gamma  \ton{1-e}^{-\gamma}\lambda^{\gamma} }{ 15 e\ton{1 - e^2}^{3/2}\ton{6-5\gamma+\gamma^2}^2 }\grf{\de\nnb
\nb\si (\gamma  (\gamma  (16 \gamma  (8 \gamma -61)+2687)-3204)+1405) e^5+\de\nnb
\nb +\si (\gamma  (\gamma  (64 \gamma -447)+939)-590) e^4-\de\nnb
\nb -\si 2 (\gamma  (\gamma  (120 \gamma -649)+1066)-525) e^3-\de\nnb
\nb -\si 10 (3\gamma -8) (7 \gamma -11) e^2+\de\nnb
\nb +\si 5 (3 (\gamma -8) \gamma +37) e-\de\nnb
\nb -\si 15 (\gamma -3) (\gamma -2)\de\nnb
& \si}\ _2 F_1\ton{\rp{3}{2},1+\gamma;2;\rp{2e}{-1+e}},\nnb
p_{16} \lb{p16} \nb = -\rp{  \gamma  \ton{1-e}^{-\gamma}\lambda^{\gamma} }{ 15 \ton{1 - e}^{3/2}\ton{1 + e}^{5/2}\ton{6-5\gamma+\gamma^2}^2 }\grf{\de\nnb
\nb\si 2 (\gamma -2) (\gamma  (2 \gamma  (8 \gamma  (8 \gamma -55)+1093)-2421)+985) e^5+\de\nnb
\nb +\si (\gamma  (2 \gamma  (24 \gamma  (4 \gamma -29)+1793)-3925)+1585) e^4-\de\nnb
\nb -\si 4 (\gamma  (3 \gamma  (10\gamma  (4 \gamma -29)+741)-2402)+955) e^3+\de\nnb
\nb +\si (2290-2 \gamma  (2 \gamma  (60 \gamma -413)+1759)) e^2+\de\nnb
\nb +\si 10 (\gamma  (3 \gamma  (6 \gamma -31)+137)-52) e+\de\nnb
\nb +\si 5 (9 \gamma  (2 \gamma-9)+89)\de\nnb
& \si}\ _2 F_1\ton{\rp{3}{2},1+\gamma;3;\rp{2e}{-1+e}}.
\end{align}

\bibliography{EPHEMERIDESbib}{}
\bibliographystyle{mdpi-arXiv}

\end{document}